\documentclass[10pt, conference]{IEEEtran}
 \IEEEoverridecommandlockouts

\usepackage{cite}
\usepackage{amsmath,amssymb,amsfonts}
\usepackage{graphicx}
\usepackage{textcomp}
\usepackage{xcolor}
\usepackage{acro}
\usepackage{xurl}
\usepackage{siunitx}
\usepackage{booktabs}
\usepackage{tabularx}
\usepackage{multirow}
\usepackage{algorithmicx}
\usepackage{algorithm}
\usepackage[noend]{algpseudocode}
\usepackage{mathtools}
\usepackage{graphicx}
\usepackage{subcaption}

\usepackage[left=0.63in, right=0.631in,top=1.9cm, bottom=2.67cm]{geometry}

\usepackage{scalerel}
\usepackage{tikz}
\usepackage{pgfplots}
\pgfplotsset{compat=1.18}
\usetikzlibrary{svg.path}
\definecolor{orcidlogocol}{HTML}{A6CE39}
\tikzset{
  orcidlogo/.pic={
    \fill[orcidlogocol] svg{M256,128c0,70.7-57.3,128-128,128C57.3,256,0,198.7,0,128C0,57.3,57.3,0,128,0C198.7,0,256,57.3,256,128z};
    \fill[white] svg{M86.3,186.2H70.9V79.1h15.4v48.4V186.2z}
                 svg{M108.9,79.1h41.6c39.6,0,57,28.3,57,53.6c0,27.5-21.5,53.6-56.8,53.6h-41.8V79.1z M124.3,172.4h24.5c34.9,0,42.9-26.5,42.9-39.7c0-21.5-13.7-39.7-43.7-39.7h-23.7V172.4z}
                 svg{M88.7,56.8c0,5.5-4.5,10.1-10.1,10.1c-5.6,0-10.1-4.6-10.1-10.1c0-5.6,4.5-10.1,10.1-10.1C84.2,46.7,88.7,51.3,88.7,56.8z};
  }
}
\newcommand\orcidicon[1]{\href{https://orcid.org/#1}{%
\mbox{\begin{tikzpicture}[yscale=-1,transform shape, scale=0.03] 
\pic{orcidlogo};
\end{tikzpicture}}}}

\usepackage{hyperref}
\hypersetup{
    colorlinks = true,
    allcolors=blue,
    urlcolor=black,
    linkcolor=black
}

\usepackage{tablefootnote}

\usepackage{xspace}
\let\OldTexttrademark\texttrademark
\renewcommand{\texttrademark}{\OldTexttrademark\xspace}%

\DeclareSIUnit{\kbps}{k\acl{bit}/s}
\DeclareSIUnit{\mbps}{M\acl{bit}/s}
\DeclareSIUnit{\gbps}{G\acl{bit}/s}
\DeclareSIUnit{\tbps}{T\acl{bit}/s}
\DeclareSIUnit{\B}{\acl{byte}}
\DeclareSIUnit{\kB}{k\acl{byte}}
\DeclareSIUnit{\mB}{M\acl{byte}}
\DeclareSIUnit{\gB}{G\acl{byte}}
\DeclareSIUnit{\tB}{T\acl{byte}}
\DeclareSIUnit{\b}{\acl{bit}}
\DeclareSIUnit{\kb}{k\acl{bit}}
\DeclareSIUnit{\mb}{M\acl{bit}}
\DeclareSIUnit{\gb}{G\acl{bit}}
\DeclareSIUnit{\tb}{T\acl{bit}}
\DeclareSIUnit{\mpps}{Mpps}

\newcolumntype{Y}{>{\centering\arraybackslash}X}

\newcommand*\ballnumber[1]{\tikz[baseline=(char.base)]{
            \node[shape=circle,fill,inner sep=.5pt] (char) {\textcolor{white}{#1}};}}

\def\BibTeX{{\rm B\kern-.05em{\sc i\kern-.025em b}\kern-.08em
    T\kern-.1667em\lower.7ex\hbox{E}\kern-.125emX}}

\usepackage{fancyhdr}
\fancypagestyle{IEEEtitlepagestyle}{
    \fancyhf{} 
     
    \fancyhead[C]{
        \small © 2026 IEEE. Personal use of this material is permitted. Permission from IEEE must be obtained for all other uses, including reprinting/republishing this material for advertising or promotional purposes, collecting new collected works for resale or redistribution to servers or lists, or reuse of any copyrighted component of this work in other works.\\
        \small \textit{Digital Object Identifier 10.1109/NetSoft70012.2026.11603500}
        }
}

\newcommand\citeN\cite

\newcommand\fig[1]{Figure~\ref{fig:#1}}

\newcommand\sect[1]{Section~\ref{sec:#1}}
\newcommand\equa[1]{Equation~\ref{eq:#1}}
\newcommand\tabl[1]{Table~\ref{tab:#1}}

\newcommand{\figeps}[3][]{%
  \begin{figure}[t]
    \centering
    \captionsetup{aboveskip=1pt, belowskip=-2pt}
    \includegraphics[width=#1]{figures/#2}
    \caption{#3}
    \label{fig:#2}
  \end{figure}
}

\newboolean{makevspace}
\newcommand{\cvspace}[1]{%
   \ifthenelse
   {\boolean{makevspace}}
   {\vspace{#1}}
   {}%
}
\acsetup{single}

\DeclareAcronym{TG}{
    short = TG,
    long = traffic generator,
}

\DeclareAcronym{DDoS}{
    short = DDoS,
    long = distributed denial-of-service,
}

\DeclareAcronym{RTT}{
    short = RTT,
    long = round-trip time,
}

\DeclareAcronym{DuT}{
    short = DuT,
    long = device under test,
    short-plural = DuTs,
    plural = devices under test,
}

\DeclareAcronym{IAT}{
    short = IAT,
    long = inter-arrival time,
}

\DeclareAcronym{MAT}{
    short = MAT,
    long = match-action table,
}

\DeclareAcronym{CBR}{
    short = CBR,
    long = constant bit rate,
}

\DeclareAcronym{TCAM}{
    short = TCAM,
    long = ternary content addressable memory,
}

\DeclareAcronym{byte}{
    short = B,
    long = byte,
}

\DeclareAcronym{bit}{
    short = b,
    long = bit,
}

\DeclareAcronym{NRMSD}{
    short = NRMSD,
    long = normalized root-mean-square deviation,
}

\begin{document}

\title{High-Speed Generation of Periodic Traffic Patterns on P4TG for DDoS and Burst-Load Evaluation}

\author{
\IEEEauthorblockN{Fabian Ihle,
Etienne Zink,
and Michael Menth}
\IEEEauthorblockA{University of T\"ubingen, Chair of Communication Networks
\\Email: \{fabian.ihle,
etienne.zink, 
menth\}@uni-tuebingen.de}
  \thanks{
  The authors acknowledge the funding by the Deutsche Forschungsgemeinschaft (DFG) under grant 503231190, and the use of Claude Sonnet 4.5 to assist in developing scripts for experiment automation.}
  }

\maketitle

\begin{abstract}
Traffic generators are essential tools for evaluating the robustness and performance of networked systems.
P4TG is an open-source, hardware-accelerated traffic generator implemented in P4 for the Intel Tofino™ ASIC.
It has been adopted by researchers and industry due to its flexibility and multi-terabit generation capability, and its low cost compared to other traffic generators.
However, like most existing generators, it primarily produces constant bit rate traffic, which does not reflect the highly time-varying behavior observed in real networks, such as flashcrowds and microbursts.
Such patterns are difficult to emulate at scale with current tools.
We present a data plane mechanism for P4TG that shapes periodic, time-varying traffic patterns, including patterns representative of DDoS attacks and burst-load scenarios.
Pattern shaping in P4TG can be applied to its generated traffic at an aggregate throughput of up to 4 Tbit/s.
We evaluate pattern accuracy and analyze scalability across different sampling resolutions and periods.
Further, we demonstrate practical use cases, including zero-loss throughput determination and buffer capacity measurement.
Finally, we present microburst-based attack scenarios that overload UDP receivers, switch buffers, and degrade TCP throughput on shared links while remaining undetectable to conventional rate monitoring.
\end{abstract}


\vspace{-0.3cm}

\section{Introduction}
\label{sec:introduction}
Traffic generators are essential for evaluating the performance and resilience of networked systems.
They enable controlled experiments by producing and analyzing packet streams that stress devices and protocols under well-defined conditions.
Software-based generators are flexible and inexpensive, but cannot sustain the line rate performance required for modern high-speed networks.
Hardware generators offer much higher throughput, yet at the cost of a lack of programmability and significantly higher expenses.
This gap motivates the development of programmable, hardware-accelerated generators that combine flexibility with terabit-scale performance.

P4TG~\cite{LiHae23} is a recent example of such a system.
Implemented in P4 on Intel Tofino\texttrademark\ ASICs, it provides high configurability and supports an aggregate traffic generation capacity of up to \qty{4}{\tbps}.
While P4TG currently generates \ac{CBR} and Poisson-distributed traffic, which are well suited for evaluating sustained load conditions, real network traffic often deviates from such rates.
In many scenarios, traffic shows highly variable and time-dependent patterns.
Flashcrowds, for example, occur when large numbers of legitimate users simultaneously access a service, creating rapid spike, plateau, and decay phases~\cite{BeKu17,ArHo03,SiSi22}.
Malicious workloads such as \ac{DDoS} attacks similarly show characteristic patterns, including pulsing or on/off behavior~\cite{BrBr17,AsUs25,LiWu24,JiMi05,ShRe20}.
Short-lived microbursts, i.e., high-rate spikes lasting only microseconds, have been widely observed in operational networks and are known to cause packet loss while often remaining invisible to coarse-grained measurements~\cite{ZhLi17,JoQu18,ShRe20}.
For controlled evaluation, these patterns can be approximated as periodic functions with configurable amplitude and frequency.
Accurately reproducing them is crucial for evaluating how networks react to overload conditions, yet existing traffic generators do not support patterns at multi-terabit rates~\cite{AlSi23, EmGa15, TRex,KuSi20,CoVo24}.

The contribution of this work is manifold.
We introduce a programmable traffic pattern abstraction for P4TG that enables shaping of periodic traffic patterns at line rate.
Our design performs pattern shaping entirely in the data plane while the control plane configures the pattern.
This enables P4TG to reproduce a wide range of time-varying patterns, defined through mathematical functions at an aggregate throughput of up to \qty{4}{\tbps}.
We provide exemplary pattern functions, including sine, square, sawtooth, and flashcrowd.
We analyze the scalability of the mechanism in terms of maximum period and table footprint and show that the generated traffic accurately follows the configured patterns.
We show its applicability with practical use cases that determine the zero-loss throughput, and measure the buffer capacity of a device.
Further, we demonstrate three microburst-based attack scenarios which overload a UDP target, a switch buffer, and a shared TCP link while remaining nearly invisible to conventional monitoring systems.
These capabilities allow controlled, high-speed evaluation of networked systems under realistic load conditions that are difficult to reproduce with existing traffic generators.
\section{Technical Background}
\label{sec:background}
In this section, we provide a brief introduction to the P4 programming language and the Intel Tofino\texttrademark\ ASIC, followed by an overview of the traffic generator P4TG.

\subsection{The P4 Programming Language and the Intel Tofino}
P4~\cite{BoDa14} is a domain-specific programming language for the data plane of network devices.
It defines how packets are parsed and processed in a pipelined structure.
A P4 program consists of a programmable parser, a sequence of control blocks, and a programmable deparser.
Each control block contains \acfp{MAT} that select actions based on packet header fields or metadata.
Packets are matched against these tables using pre-defined key fields, and the associated action is then executed.
\acp{MAT} support several match types, including exact, ternary, and range matches.

The Intel Tofino\texttrademark\ 1 and 2 switching ASICs~\cite{PublicTNA} are hardware platforms that execute P4 programs at line rates of up to \qty{400}{\gbps} per port.
They follow a pipelined architecture in which each stage can perform table lookups and simple arithmetic operations.
Extern objects extend the functionality of P4.
Meter externs implement a token-bucket algorithm and are used for rate control and shaping. 
Register externs provide stateful memory that allows P4 programs to maintain state across packets.
In addition, the Intel Tofino\texttrademark\ provides a configurable packet generation extern that can inject packets with up to \qty{400}{\gbps} into the pipeline from an internal generator.

\subsection{The Traffic Generator P4TG}
P4TG~\cite{LiHae23} is an open-source~\cite{p4tg-git} traffic generator implemented in the P4 programming language for the Intel Tofino\texttrademark\ 1 and 2 switching ASICs.
It leverages the ASIC's internal traffic generation port to produce high-speed packet streams that can be internally multicasted to up to ten physical ports.
This enables an aggregate generation capacity of up to \qty{4}{\tbps} ($10\times \qty{400}{\gbps}$) on the Intel Tofino\texttrademark\ 2.
P4TG supports configurable traffic based on Ethernet, IPv4, and IPv6, and can optionally apply a variety of encapsulations, including VLAN, QinQ, MPLS, VxLAN, and SRv6~\cite{IhZi25}.
The frame size, header fields, and traffic characteristics, such as the generation rate, can be freely configured, allowing users to generate \ac{CBR} as well as Poisson-distributed traffic.
Further, P4TG supports IPv4/6 address randomization enabling millions of different IP flows, e.g., for emulating sources of \ac{DDoS} attacks.
Generated packets carry a lightweight UDP header that embeds P4TG-specific metadata, such as sequence numbers and timestamps, to facilitate detailed performance analysis.
The data plane collects extensive traffic statistics, including \acp{RTT} and \acp{IAT} (optionally via histogram-based measurement~\cite{IhZi25_2}), TX/RX rates, packet loss, and detailed frame type and size distributions.
These statistics are exported to the control plane through a REST API and visualized in real time using a web-based frontend.
P4TG has been adopted by various research groups and industrial stakeholders, including Airbus~\cite{MaKh25} and the German Federal Government~\cite{bdbos}, to evaluate network prototypes and operational environments.

While P4TG provides comprehensive generation and measurement capabilities, it does not perform traffic pattern shaping beyond \ac{CBR} or Poisson-distributed traffic.
Therefore, in this work, we introduce periodic pattern shaping for generated traffic in P4TG.

\section{Related Work}
\label{sec:rel_work}
In this section, we first review related work on traffic patterns that characterize DDoS attacks and flashcrowds.
Next, we summarize the generation of periodic traffic patterns in other traffic generators.

\subsection{Traffic Patterns and DDoS Characteristics}
We first summarize traffic patterns, and then characterize DDoS traffic.
\subsubsection{Traffic Patterns}
\label{sec:rel_work_patterns}
Traffic on the Internet is rarely \ac{CBR}-shaped.
Instead, it shows periodicity and sudden spikes in volume.
We consider three types of traffic patterns: square waves, microbursts, and flashcrowds.

\paragraph*{Square Waves}
A number of works demonstrate that adversaries deliberately employ periodic or on/off attack patterns to evade or manipulate mitigation systems.
These pulsing on/off attack strategies can be abstracted as square-wave traffic patterns, characterized by alternating phases of high and low offered load.
Bremler-Barr \textit{et al.}~\cite{BrBr17} and Asudi \textit{et al.}~\cite{AsUs25} describe the Yo-Yo attack in which an attacker injects periodic overload bursts to force cloud auto-scaling systems to oscillate between scale-up and scale-down phases.
Li \textit{et al.}~\cite{LiWu24} analyze DNS-based pulsing attacks in which low-rate queries lead to amplified responses, producing high-volume bursts.
A recent study by Kopp \textit{et al.}~\cite{KoHo25} observed pulse-wave \ac{DDoS} attack patterns at an Internet exchange point.
They state that pulsed DDoS attacks constitute \qty{27}{\percent} of \ac{DDoS} attacks.

\paragraph*{Microbursts}
Microbursts are short-lived spikes in traffic that typically last only tens to hundreds of microseconds~\cite{ZhLi17}, yet they can cause sharp increases in latency and momentary packet loss~\cite{ShRe20}.
Microbursts can be viewed as square waves with a very short high phase.
Microbursts may originate from queuing inside network devices~\cite{JoQu18} or may be triggered as part of \ac{DDoS} attacks.
Traditional monitoring approaches, such as SNMP~\cite{rfc3410} and NetFlow~\cite{rfc3176}, sample traffic at second-scale intervals, and therefore cannot detect such short-lived spikes.
With coarse sampling, the instantaneous burst is averaged out, the measured rate remains low, and only the resulting packet loss reveals that congestion occurred.
Consequently, several works~\cite{ZhLi17,JoQu18,GaLi23,ShRe20, ZhHu24} investigate the observability of microbursts and devise methods to detect them at microsecond- to millisecond-level resolution.
However, existing studies typically evaluate microburst behavior at modest rates of up to \qty{10}{\gbps}~\cite{JoQu18,ShRe20} or rely on observing naturally occurring bursts rather than generating them in a controlled manner.
In contrast, this work enables the generation of microburst patterns at aggregate rates of up to \qty{4}{\tbps}, providing a controlled environment to study network behavior under short-lived overload events.

\paragraph*{Flashcrowds}
The flashcrowd pattern represents a class of highly non-stationary traffic originating from legitimate users.
These events occur when a large number of users begin accessing the same service simultaneously, leading to a rapid increase in demand, e.g., during major news events, software releases, or live-streamed events.
Behal \textit{et al.}~\cite{BeKu17} show that such events can overload bandwidth, CPU, and memory resources in a manner similar to \ac{DDoS} attacks.
Ari \textit{et al.}~\cite{ArHo03} characterize flashcrowd patterns with a ramp-up phase with linear growth, a sustained high-demand plateau, and a ramp-down phase with logarithmic decay.
Flashcrowds are hard to distinguish from \ac{DDoS} attacks which is why Silva \textit{et al.}~\cite{SiSi22} propose models to distinguish them.

\subsubsection{DDoS Traffic Characteristics}
\label{sec:rel_work_ddos}
Several measurement studies quantify the temporal and volumetric characteristics of \ac{DDoS} attacks.
Mao \textit{et al.}~\cite{MaMo06} show that most \ac{DDoS} attacks last less than one hour, occasionally up to 12 hours, and can reach rates near one million packets per second (pps).
Although such rates are manageable for ISP backbones, they overwhelm servers and security appliances.
Yuan \textit{et al.}~\cite{JiMi05} and Kopp \textit{et al.}~\cite{KoHo25} report periodic \ac{DDoS} flooding with pulse durations of \qty{300}{s} and \qty{1}{minute}, respectively, forming square-wave–like on/off patterns.
Recent industry reports highlight a continued rise in hyper-volumetric attacks~\cite{cloudflare2,cloudflare3,microsoft}, i.e., attacks that exceed a rate of \qty{1}{\tbps}.
Cloudflare~\cite{cloudflare2, cloudflare3} noted a substantial increase in attacks exceeding \qty{1}{\tbps} with a record peak of \qty{29.7}{\tbps} in Q3 2025 and surpassing the previous quarter’s peak of \qty{7.3}{\tbps}.
Microsoft similarly reported a \qty{15.72}{\tbps} attack targeting an Azure endpoint~\cite{microsoft} in October 2025.
These events typically last between tens of seconds and several minutes~\cite{cloudflare2,cloudflare3}.

\subsection{Traffic Patterns in other Traffic Generators}
In the following, we discuss multiple software- and hardware-based traffic generators.
They are summarized in \tabl{tg_overview}.

\begin{table}[htb]

  \caption{Overview of traffic generators and their pattern shaping capabilities.}
  \label{tab:tg_overview}
  \setlength{\tabcolsep}{6pt}
  \renewcommand{\arraystretch}{1.05}
  \renewcommand{\tabularxcolumn}[1]{m{#1}}   
  \small
  \begin{tabularx}{\columnwidth}{|Y|Y|Y|Y|}
    \hline
    \textbf{Traffic generator} &
    \textbf{Type} &
    \textbf{Max. generation rate} &
    \textbf{Pattern shaping} \\ \hline\hline
    WAVE~\cite{AlSi23} &
    Software &
    n/a &
    Function-based \\ \hline
    MoonGen~\cite{EmGa15} &
    Software &
    \mbox{$\le \qty{120}{\gbps}$} &       
    Manual \\ \hline
    TRex~\cite{TRex} &
    Software &
    \mbox{$\le \qty{200}{\gbps}$} &
    Not supported \\ \hline
    P4STA~\cite{KuSi20} &
    Hybrid &
    None &
    Not supported  \\ \hline
    PIPO-TG~\cite{CoVo24} &
    Hardware &
    \mbox{$\le$ \qty{1}{\tbps}} &
    Manual \\ \hline
    P4TG &
    Hardware &
    \mbox{$\le$ \qty{4}{\tbps}} &
    Function-based \\ \hline
  \end{tabularx}
\end{table}

The WAVE generator by Almeida \textit{et al.}~\cite{AlSi23} orchestrates the number of concurrent application instances, e.g., \texttt{iperf}, instead of directly shaping the traffic.
The number of instances over time follows predefined functions such as sinusoid and flashcrowd, from which the traffic pattern emerges indirectly.

MoonGen~\cite{EmGa15} is a scriptable DPDK-based software generator capable of packet generation up to \qty{120}{\gbps}.
It generates traffic patterns by controlling inter-packet gaps in software, but relies on NIC-specific features and the injection of invalid packets to shape timing, which fundamentally constrains the precision of its generated load.

TRex~\cite{TRex} is a DPDK-based traffic generator that supports \ac{CBR} traffic at high speeds of up to \qty{200}{\gbps}.
However, its rate control is limited to static per-stream configurations and it cannot generate time-varying patterns such as sine waves or flashcrowd profiles.

P4STA~\cite{KuSi20} focuses on high-precision timestamping and load aggregation using P4-programmable hardware.
It integrates existing software generators such as iPerf3 and MoonGen while its P4-based ``Stamper'' can shape the outgoing rate.
However, P4STA itself does not generate traffic.
It relies entirely on external software load generators, and its shaping capabilities are limited to simple rate limiting rather than programmable, time-varying patterns.

PIPO-TG~\cite{CoVo24}, implemented on an Intel Tofino\texttrademark\ 1 ASIC, can generate traffic up to \qty{1}{\tbps} and supports customizable headers, packet sizes, and protocol fields.
PIPO-TG generates periodic traffic patterns by manually configuring a sequence of rate levels that the generator applies over time.
This approach is suitable for simple step-wise patterns.
However, since all rate levels must be specified explicitly, generating smooth patterns, such as sine waves and exponential decays as seen in flashcrowds, requires the manual definition of numerous intermediate rate steps.
This approach becomes infeasible for approximating continuous functions with sufficient fidelity.

Overall, existing traffic generators either achieve high throughput with limited pattern flexibility (TRex, PIPO-TG) or support flexible patterns at limited rates (MoonGen).
In contrast, our extension to P4TG introduces a programmable abstraction where users specify mathematical pattern functions.
Those are automatically sampled, normalized, and translated into data plane rate profiles without manual rate-step configuration.
This enables fine-grained generation, such as sinusoidal, square wave, sawtooth, flashcrowd, and microburst patterns at aggregate rates up to \qty{4}{\tbps}.

\section{Pattern Shaping Mechanism in P4TG}
\label{sec:implementation}
\begin{figure}[t]
\centering
\includegraphics[width=\columnwidth]{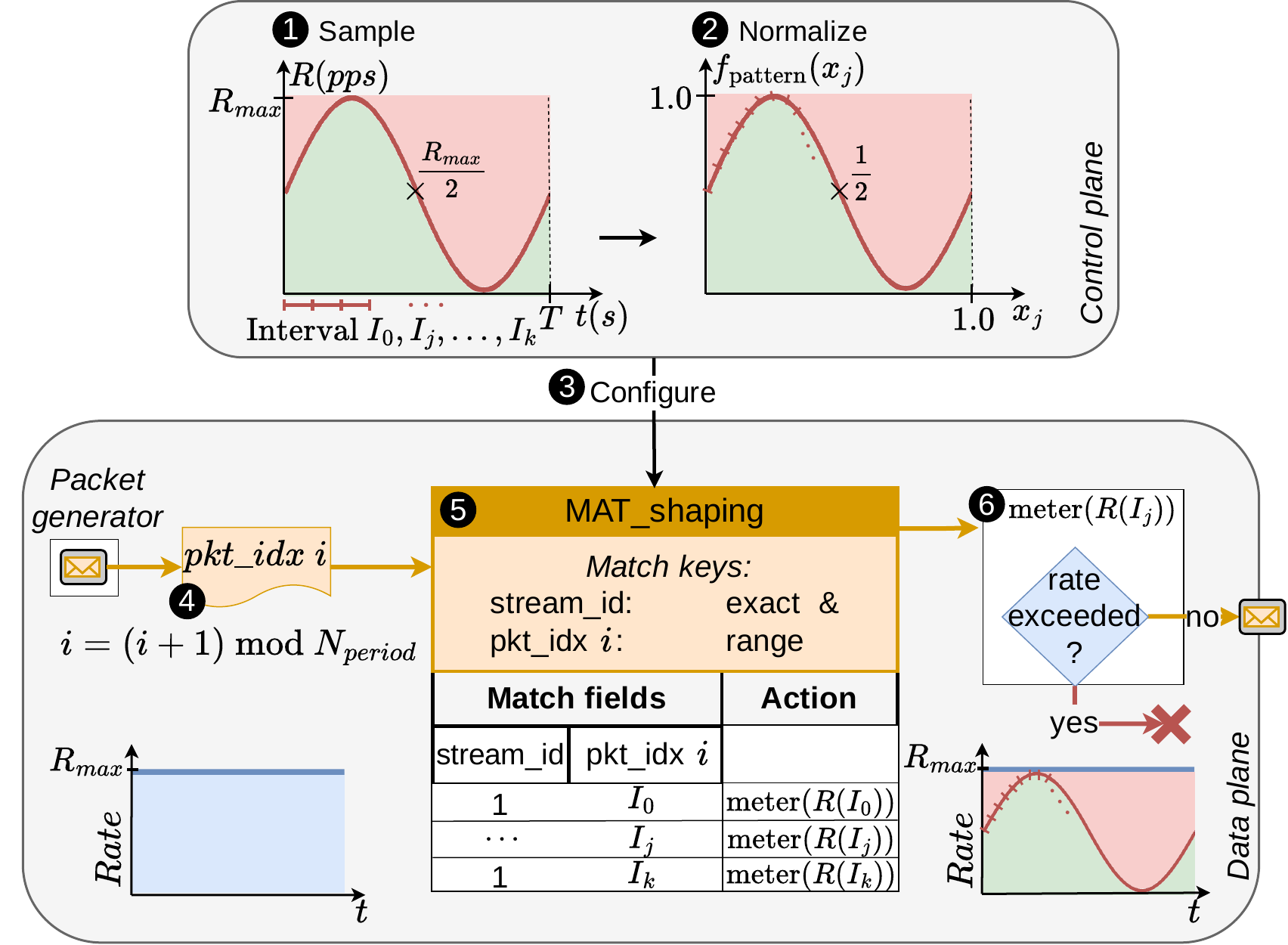}
    \captionsetup{aboveskip=3pt, belowskip=-4pt}
\caption{The control and data plane mechanisms for pattern shaping in P4TG.
}
\label{fig:pdfs/implementation-all}
\end{figure}

In this section, we describe the implementation of the pattern shaping mechanism.
Pattern shaping in P4TG is implemented through two coordinated components in the control and data plane.
They are illustrated in \fig{pdfs/implementation-all}.


First, the control plane samples the pattern, normalizes it, and installs the resulting parameters into the data plane prior to generation.
Next, the data plane determines the position of each generated packet within a period, and enforces the desired traffic pattern by dropping packets that exceed the target rate.
In the following, we describe both mechanisms in more detail.

\subsection{Sampling and Normalization of Periodic Functions}
\label{sec:impl_control_plane}
First, the user configures the desired pattern via the REST API or web-based frontend with the pattern function $f_{\text{pattern}}$, the period length $T$, the sampling factor $k$, and the maximum packet rate $R_{\text{max}}$ in packets per second (pps), i.e., the amplitude.
Then, before traffic generation starts, the control plane derives all parameters required for periodic pattern shaping which is illustrated in step \ballnumber{1} to \ballnumber{3} in \fig{pdfs/implementation-all}.
These parameters include the number of packets per period, the sampling intervals based on the sampling factor $k$, and the rate for each sampled interval.
This section covers the operations the control plane performs to sample and normalize the configured patterns.

\subsubsection{Determining the Number of Packets per Period}
Given the period $T$ and the rate $R_{\text{max}}$, the control plane first computes how many packets $N_{\text{period}}$ are generated during one full period:
\begin{align}
    N_{\text{period}} &= R_{\text{max}}\cdot T.
\end{align}
This value is given to the data plane.

\subsubsection{Sampling the Period into k Intervals}
To approximate the pattern, the period is divided into $k$ equally sized sampling intervals.
This is shown in step \ballnumber{1} in \fig{pdfs/implementation-all}.
Each interval contains
\begin{align}
    N_{\text{interval}} &= \frac{N_{\text{period}}}{k}
\end{align}
packets.
This yields the interval boundaries
\begin{align}
    I_j &= [j\cdot N_{\text{interval}},\ (j+1)\cdot N_{\text{interval}}), \quad\quad j = 0, 1, \ldots k-1.
\end{align}
These intervals define the entries of the shaping \ac{MAT} used in the data plane.
A sampling factor of $k=32$ provides a good tradeoff between the accuracy of the sampled pattern and required table space.
This is further evaluated in \sect{eval_sampling}.

\subsubsection{Mapping Intervals onto the Pattern Function}
Each interval index $j$ is normalized to a value $x_j \in[0,1)$, as illustrated in step \ballnumber{2} of \fig{pdfs/implementation-all}:
\begin{align}
    x_j &= \frac{j}{k}.
\end{align}
The normalized value is then passed to the configured pattern function
\begin{align}
    f_{\text{pattern}} &: [0,1) \times P \rightarrow [0,1],
\end{align}
which defines the desired pattern, e.g., sine, square, sawtooth, or flashcrowd.
Here, $P$ denotes the set of pattern-specific parameters, allowing functions to be parameterized, e.g., by an exponential decay rate in the flashcrowd pattern.
Examples for $f_{\text{pattern}}$ are given in \sect{patterns}.

\subsubsection{Computing the Target Rate for Each Interval}
Finally, the output amplitude of the pattern during interval $I_j$ determines the applied shaping rate:
\begin{align}
    R(I_j) &= f_{\text{pattern}}(x_j) \cdot R_{\text{max}}.
\end{align}
Afterwards, these rates are written into the shaping \ac{MAT} and configure meter externs in the data plane, shown in step \ballnumber{3} in \fig{pdfs/implementation-all}.
Traffic exceeding these rates is dropped in the data plane.

\subsection{Data Plane Enforcement of Traffic Patterns}
\label{sec:impl_data_plane}
An overview of the data plane logic is shown in step \ballnumber{4} to \ballnumber{6} in \fig{pdfs/implementation-all}.
For periodic pattern shaping, traffic is generated per stream as \ac{CBR} traffic with a configurable rate $R_{\text{max}}$.
The P4 pipeline then enforces the configured pattern by selectively dropping packets whose rate exceeds the sampled rate $R(I_j)$ for each interval (computed in \sect{impl_control_plane}).
The data plane mechanism consists of two building blocks: a packet-based mechanism for determining the position of a packet within the period, and a token-bucket algorithm for shaping.

\subsubsection{Packet-based Periodicity}
Packets need to be mapped to a position in the periodic pattern.
To determine a packet’s position within the period, P4TG maintains a per-stream packet counter in a \qty{32}{bit} register.
This counter increments with each packet and wraps around after a configured number $N_{\text{period}}$ producing a periodic packet index
\begin{align}
    i \in[0,\ N_{\text{period}}),
\end{align}
shown in step \ballnumber{4} in \fig{pdfs/implementation-all}.
This provides a lightweight approach to periodicity, though it can also produce non-periodic patterns by limiting execution to a single period.
In principle, periodicity can also be derived from timestamps~\cite{IhLi24}, which supports arbitrarily sized intervals but requires significant pipeline resources.
Since P4TG discretizes patterns into $k$ equally sized intervals, a wrap-around packet counter is sufficient and more resource-efficient.
 
\subsubsection{Pattern Shaping}
Once the position of a packet within a period is determined, the packet must be metered according to the pattern.
For this purpose, the pipeline contains a \ac{MAT} that stores $k$ equally-sized intervals of one period, i.e., the intervals $I_j$ calculated in \sect{impl_control_plane}.
Based on the current packet's position within the period, it matches one of these intervals in the \ac{MAT}, shown in step \ballnumber{5}.
Each interval is associated with a meter extern, implementing a token bucket algorithm as defined in RFC 2698~\cite{RFC2698}.
Here, tokens accumulate at a configured rate, and each packet consumes tokens proportionally to its size.
Packets for which no tokens remain are dropped.
The token generation rate for each interval directly corresponds to the target rate $R(I_j)$ computed by the control plane.
Packets exceeding $R(I_j)$ are therefore dropped, shown in step \ballnumber{6}.
With this approach, shaping is executed entirely in the data plane without requiring dynamic reconfiguration of the traffic generator, enabling pattern shaping at line rate.

\subsubsection{Range-to-Ternary Conversion}
\label{sec:impl_range_ternary}
In the data plane, packets are matched to the sampled intervals of the shaping \ac{MAT} based on the \qty{32}{bit} packet index within the current period.
Although P4 supports native range matching, i.e., matching a field to an interval, the Intel Tofino\texttrademark\ architecture restricts range matches to fields of at most \qty{20}{bit}.
As a result, the full \qty{32}{bit} packet index cannot be matched using hardware-supported range operators.
To overcome this limitation, we employ a range-to-ternary conversion algorithm~\cite{GuMc01}.
The algorithm decomposes a single numeric interval with a lower and an upper bound into a set of multiple ternary (wildcard) entries whose union covers the original interval.
This enables efficient matching over the entire \qty{32}{bit} space at the cost of additional \ac{MAT} entries.
The resulting increase in table footprint and its implications for pattern shaping are evaluated in \sect{eval_range_ternary}.

\subsection{Pattern Functions for Traffic Shaping}
\label{sec:patterns}

P4TG supports shaping functions $f_{\text{pattern}}$ defined over the normalized interval $[0,\ 1)$.
In the current version, the control plane provides four built-in functions for traffic patterns: sine, square, sawtooth, and flashcrowd.
New functions can be easily added to the control plane.
Each function maps a normalized phase value $x \in [0,1)$ to an amplitude in $[0,1]$ which directly scales the target rate for the corresponding sampling interval.
The function and all its parameters, including the period and sampling factor, are fully integrated into the web-based frontend and REST API for configuration.
The definitions of these functions are given below.

\noindent
\paragraph*{Sine Pattern}
The sine pattern is a smooth periodic sine wave normalized to $[0, 1]$:
    \begin{equation}
    \small
        f_{\text{sine}}(x)=\tfrac{1}{2}\bigl(1+\sin(2\pi x)\bigr).
    \end{equation}
    
    


While sinusoidal patterns are not explicitly discussed in prior work on traffic characterization, we include a sine function to demonstrate the ability to model continuous functions.
It also serves as a convenient reference pattern for evaluating shaping accuracy in \sect{eval_accuracy}.

\noindent
\paragraph*{Square Wave Pattern}
The square wave pattern is a binary on/off pattern with high and low phases where the rate during the low phase is defined by the $low$ parameter, and the high phase lasts until $t_{\text{high}}$:
    \begin{equation}
    \small
        f_{\text{square}}(x, \text{low}, t_{\text{high}})=
        \begin{cases}
            1, & 0 \le x < t_{\text{high}},\\[4pt]
            \text{low}, & t_{\text{high}} \le x < 1.
        \end{cases}
    \end{equation}
    
    
    
    


This pattern models pulsing or on/off \ac{DDoS} attacks, and microbursts such as those described in \sect{rel_work_patterns}.
Three microburst-based attack scenarios using this pattern are evaluated in \sect{eval_microburst} to \sect{eval_tcp}.




\noindent
\paragraph*{Sawtooth Pattern}
The sawtooth pattern contains a continuous linear increase with a reset at the period boundary:
    \begin{equation}
    \small
        f_{\text{sawtooth}}(x)=x.
    \end{equation}
    
    
    

It facilitates the determination of the zero-loss throughput~\cite{rfc2544} of a \ac{DuT} which is described and applied in \sect{eval_throughput}.


\noindent
\paragraph*{Flashcrowd Pattern}
The flashcrowd pattern has a rate of zero until $t_0$, followed by a linear ramp-up until $t_1$, followed by exponential decay with the decay rate $\lambda$:

\begin{equation}
\small
f_{\text{flash}}(x,t_0,t_1,\lambda)=
\begin{cases}
0, & \hspace{-6pt} 0 \le x < t_0,\\[6pt]
\dfrac{x-t_0}{(t_1-t_0)}, & \hspace{-6pt} t_0 \le x < t_1,\\[10pt]
\min\!\left(1,\;
e^{\left(-\lambda\,\dfrac{x-t_1}{1 - t_1}\right)}
\right),
& \hspace{-6pt} t_1 \le x < 1.
\end{cases}
\end{equation}

This pattern models flashcrowd events and \ac{DDoS} attacks.
\section{Evaluation}
\label{sec:eval}
In this section, we first evaluate the scalability of the proposed mechanism.
Next, we demonstrate various applications, including a buffer capacity measurement, a zero-loss throughput determination, and microburst-based attacks on a UDP target, on a switch, and on a shared TCP link.

\subsection{Scalability}
We evaluate the maximum configurable period for pattern shaping, the impact of the sampling factor $k$, the number of required entries in the pattern shaping \ac{MAT}, and the accuracy of the generated patterns at terabit rates.

\subsubsection{Maximum Period for Patterns}
\label{sec:eval_period}
For pattern periodicity, the number of packets in a period is calculated based on the packet rate as described in \sect{impl_control_plane}.
This value determines the wrap-around point of the packet index register in the data plane.
Since this register is \qty{32}{bit} wide, it can represent at most $2^{32}$ distinct packet indices.
Consequently, the maximum number of packets per period, and therefore the maximum achievable period length, is bounded by this limit.
Given a packet rate $R_{\text{max}}$ (in pps), the maximum configurable period is
\begin{align}
    T_{\text{max}} &= \frac{2^{32}}{R_{\text{max}}}.
\end{align}
\tabl{tmax_calculation} lists example values of $T_{\text{max}}$ for different packet rates, and shows that P4TG supports pattern periods that match those reported for real-world \ac{DDoS} attacks in \sect{rel_work_ddos}.

\begin{table}[htb]
  \caption{Maximum period $T_{\text{max}}$ for different link rates and frame sizes.}
  \label{tab:tmax_calculation}
  \setlength{\tabcolsep}{6pt}
  \renewcommand{\arraystretch}{1.05}
  \small

  \begin{tabularx}{\columnwidth}{|Y|Y|Y|Y|}
    \hline
    \textbf{Rate (Gbit/s)} &
    \textbf{L2 frame size (Byte)} &
    \textbf{Rate $R_{\text{max}}$ (Mpps)} &
    \textbf{Maximum period $T_{\text{max}}$ (s)} \\ \hline\hline

    100 & 1518 & 8.12 & 528 \\ \hline
    100 & 512 & 23.46 & 182 \\ \hline
    100 & 64 & 148.15 & 28 \\ \hline
    400 & 1518 & 32.4 & 132 \\ \hline
    400 & 512 & 93.75 & 45 \\ \hline
    294$^{*}$ & 64 & 437.5 & 9 \\ \hline

  \end{tabularx}\\[4pt]
  \footnotesize
  $^{*}$Maximum rate P4TG can generate with \qty{64}{B} frames on a single port.  
\end{table}


\subsubsection{Impact of the Sampling Factor}
\label{sec:eval_sampling}
The control plane samples each periodic pattern into $k$ discrete intervals.
A higher sampling factor improves the accuracy of the shaped pattern but also increases the required \ac{MAT} space.
In this experiment, we evaluate how different sampling factors $k$ affect the pattern shape.
For that purpose, we configure four \qty{100}{\gbps} streams for generation in parallel with different sampling factors $k \in \{8, 16, 32, 64\}$ and a period of $T = \qty{40}{s}$.
We use a sine-wave pattern as it most clearly exposes differences in sampling accuracy, and measure the generated TX rate with P4TG.
Traffic is generated for \qty{60}{s} and the experiment is repeated ten times.
The aggregated results are shown in \fig{pdfs/sampling_40s.pdf}.

%


%

\figeps[0.8\columnwidth]{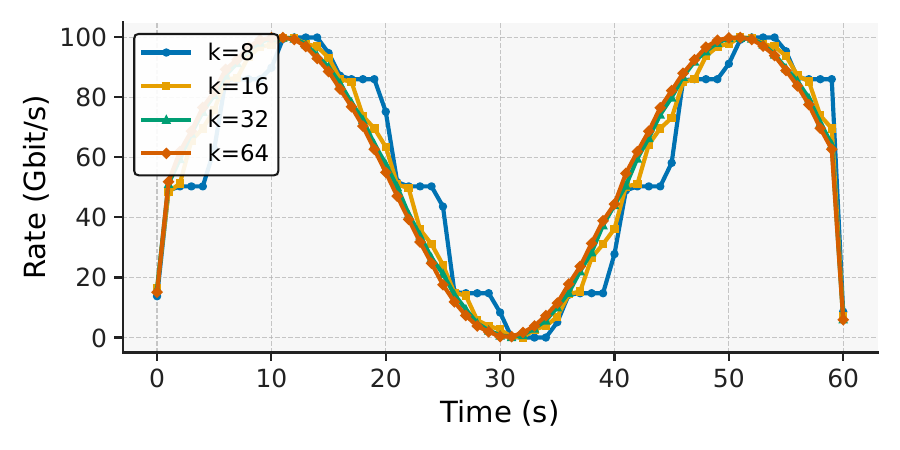}{Measured TX rate in P4TG with different sampling factors.}

\fig{pdfs/sampling_40s.pdf} illustrates that with a low sampling factor of $k \in \{8, 16\}$, the resulting waveform deviates noticeably from the ideal sine shape.
Sampling factors $k \in \{32, 64\}$ produce smooth waveforms that closely match the target function. 
Therefore, $k = 32$ serves as a practical default.
However, this value can be configured using the web frontend, or the REST API.

\subsubsection{MAT Resource Usage for Pattern Shaping}
\label{sec:eval_range_ternary}
The control plane divides the configured period into $k$ equally sized sampling intervals and installs them into the shaping \ac{MAT} of the data plane.
The required number of \ac{MAT} entries varies depending on the configured parameters, i.e., the sampling factor $k$, the traffic rate $R_{\text{max}}$, and the period $T$.
Since the Intel Tofino\texttrademark\ architecture supports range matching only for fields up to \qty{20}{bit}, interval matching over the full \qty{32}{bit} packet index requires the range-to-ternary conversion described in \sect{impl_range_ternary}.
This conversion expands each numeric interval into multiple ternary entries whose union covers the same range, thereby increasing the number of required \ac{MAT} entries.
\fig{pdfs/mat_entries_vs_rate} shows an overview of the number of required ternary \ac{MAT} entries for $R_{\text{max}} \in \{10, 50, 100, 150\}$ Mpps, $k \in \{16,32, 64, 128\}$, and $T \in \{20, 40, 60\}$ s.

\figeps[0.65\columnwidth]{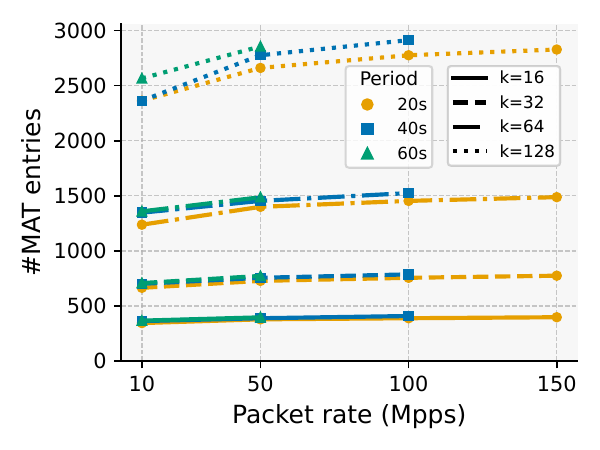}{Required number of ternary \ac{MAT} entries based on the packet rate $R_{\text{max}}$, the period $T$, and the sampling factor $k$.}

Missing data points in \fig{pdfs/mat_entries_vs_rate}, e.g., for $T=60$ and $R_{\text{max}}=\qty{150}{Mpps}$, result from the period exceeding the maximum duration as evaluated in \sect{eval_period}.
Generally, the number of required \ac{MAT} entries increases with the period length and packet rate, and ranges from a minimum of $345$ for $k=16, T=\qty{20}{\s}$ to a maximum of $2913$ for $k=128, T=\qty{40}{\s}$.
Using fewer sampling intervals significantly reduces the required number of \ac{MAT} entries.
For $k=32$, the number of required entries ranges from $665$ to $773$ while still providing sufficient shaping granularity for the evaluated patterns as shown in \sect{eval_sampling}.
Accordingly, P4TG uses $k=32$ as the default configuration.
The current shaping \ac{MAT} provides space for 40000 entries which is sufficient to accommodate multiple patterns simultaneously.

\subsubsection{Pattern-Rate Accuracy at High Throughput}
\label{sec:eval_accuracy}
\begin{figure}[t]
    \centering

    \begin{subfigure}[t]{0.49\linewidth}
        \includegraphics[width=\linewidth]{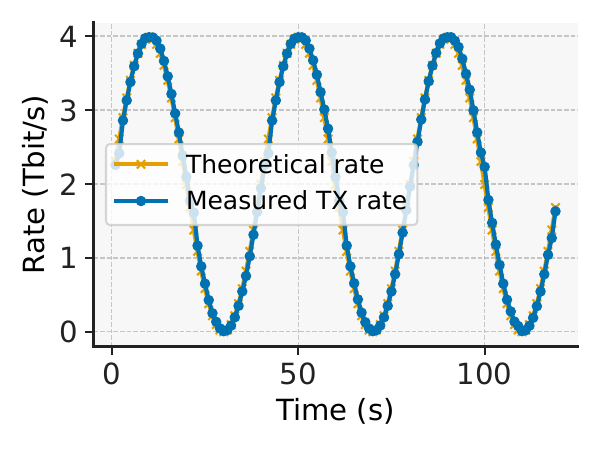}
        \caption{Sine.}
        \label{fig:sine}
    \end{subfigure}
    \hfill
    \begin{subfigure}[t]{0.49\linewidth}
        \includegraphics[width=\linewidth]{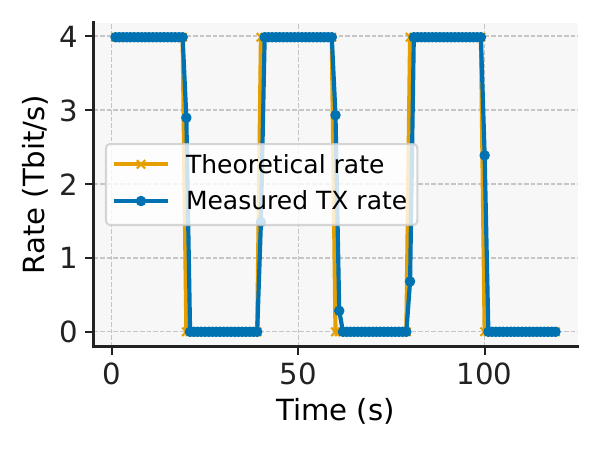}
        \caption{Square.}
        \label{fig:square}
    \end{subfigure}

    \begin{subfigure}[t]{0.49\linewidth}
        \includegraphics[width=\linewidth]{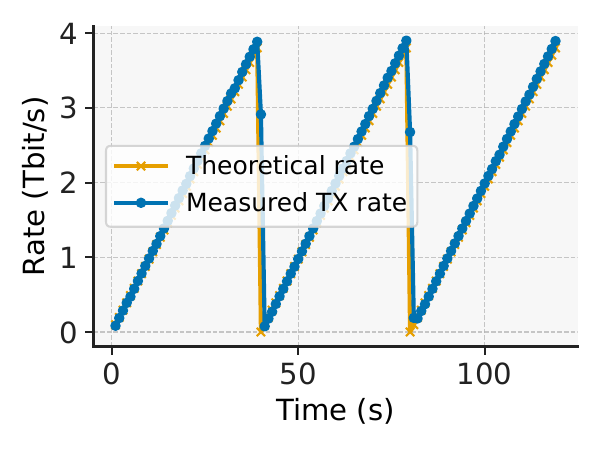}
        \caption{Sawtooth.}
        \label{fig:sawtooth}
    \end{subfigure}
    \hfill
    \begin{subfigure}[t]{0.49\linewidth}
        \includegraphics[width=\linewidth]{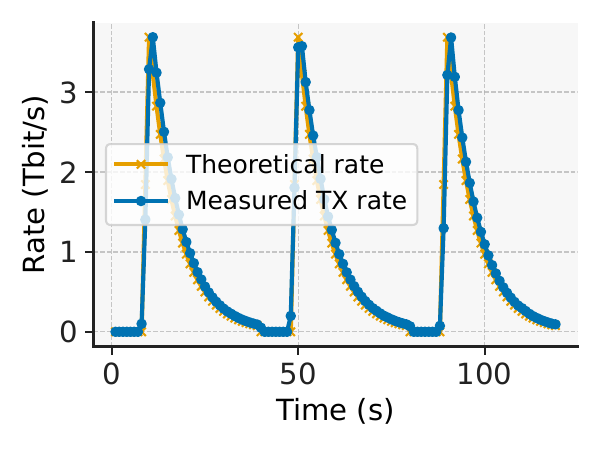}
        \caption{Flashcrowd.}
        \label{fig:flashcrowd}
    \end{subfigure}


    \caption{Generated traffic patterns measured in P4TG.}
    \label{fig:pattern_rates}
\end{figure}

In this section, we assess the accuracy of the generated pattern shapes.
We configure P4TG to generate \qty{4}{\tbps} of traffic across ten \qty{400}{\gbps} ports using \qty{1518}{\byte} frames.
This is the maximum rate P4TG can generate, but pattern shaping can be applied at any lower rate.
We evaluate four pattern functions, i.e., $f_{\text{pattern}} \in \{f_{\text{sine}}, f_{\text{square}}, f_{\text{sawtooth}}, f_{\text{flash}}\}$, each with a period of \qty{40}{s}.
For $f_{\text{flash}}$, we use the parameters $t_0 = 0.2$, $t_1=0.25$, and $\lambda=4$.
Traffic is generated for \qty{120}{\s} resulting in three full cycles per pattern, and the experiment is repeated ten times.
The traffic is looped back to P4TG, where the L1 TX and RX rates are measured.
The theoretical rates and the aggregated per-port results are shown in \fig{pattern_rates}.
As the TX and RX rates are identical for all experiments, only the TX rates are displayed for readability.

\fig{pattern_rates} illustrates that the measured L1 rates closely follow the configured patterns across all three cycles and all pattern types.
To quantify the similarity between the theoretical and observed pattern, we compute the \ac{NRMSD}, defined as
\begin{equation}
\small
\mathrm{RMSD} =
\sqrt{\frac{1}{N}\sum_{i=1}^{N}\left(y_i-\hat{y}_i\right)^2}
\end{equation}

\begin{equation}
\small
\mathrm{NRMSD} =
\frac{\mathrm{RMSD}}{\max(\hat{y}_i)-\min(\hat{y}_i)}
\end{equation}

An \ac{NRMSD} of 0.1 indicates that the measured values deviate from the ideal pattern by \qty{10}{\percent} of the pattern’s total rate range.
Across the evaluated patterns, the \ac{NRMSD} remains low, with
$\text{NRMSD}(f_{\text{sine}}) = 0.02$,
$\text{NRMSD}(f_{\text{square}}) = 0.14$,
$\text{NRMSD}(f_{\text{sawtooth}}) = 0.09$,
and $\text{NRMSD}(f_{\text{flash}}) = 0.04$.
For $f_{\text{square}}$, the \ac{NRMSD} is slightly higher than for the other patterns because of its abrupt transitions between minimum and maximum rates.
Because P4TG reports rates as averages over fixed \qty{1}{\s} measurement intervals, intervals that contain a rate transition reflect an average of the high- and low-rate phases, producing intermediate values at the transition points.
This smoothing effect is limited to the measurement process and does not affect the traffic generation.
Overall, the results show that P4TG accurately reproduces diverse rate profiles even at an aggregate throughput of up to \qty{4}{\tbps}, demonstrating that fine-grained pattern shaping is feasible at the highest supported line rates.

\subsection{Microburst Impact on a UDP Target}
\label{sec:eval_microburst}
In this section, we demonstrate a zero-loss throughput determination, and then analyze the behavior of a UDP target under periodic microburst traffic.

\subsubsection{Testbed}
The testbed for the microburst attack emulation is illustrated in \fig{pdfs/evaluation_setup}.
\figeps[0.5\columnwidth]{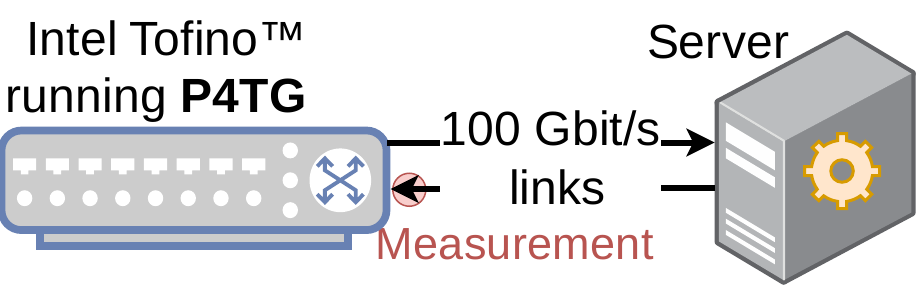}{The testbed setup.}
Traffic is generated by P4TG using \qty{1518}{B} frames and forwarded over a \qty{100}{\gbps} link to a server equipped with a ConnectX-5 NIC.
On the server, a kernel-bypass application receives the generated UDP traffic and returns all packets to P4TG for measurement.
It is not provisioned to sustain \qty{100}{\gbps} and can only process up to a limited rate before packet loss occurs.
This makes it a suitable target for testing short-lived overload events.

\subsubsection{Zero-Loss Throughput Determination}
\label{sec:eval_throughput}
We first determine the maximum rate the application can receive without loss.
RFC~2544~\cite{rfc2544} defines the zero-loss throughput $R_\text{ZLT}$ as the highest rate at which a \ac{DuT} forwards packets without packet loss.
The zero-loss throughput is determined by repeatedly adjusting the generated rate and observing packet loss.
Using pattern shaping, this procedure can be performed in a single continuous experiment.
For that purpose, P4TG generates a sawtooth pattern with a peak rate of $R_{\text{max}} = \qty{100}{\gbps}$ and a period of \qty{20}{\s}.
Generated traffic is returned to P4TG where packet loss is measured.
The experiment is repeated ten times.
The rate at which the first loss occurs defines the receiver’s saturation point $R_\text{ZLT}$.
\fig{pdfs/Baseline} shows the results of the zero-loss throughput measurement using the sawtooth pattern.

\figeps[0.8\columnwidth]{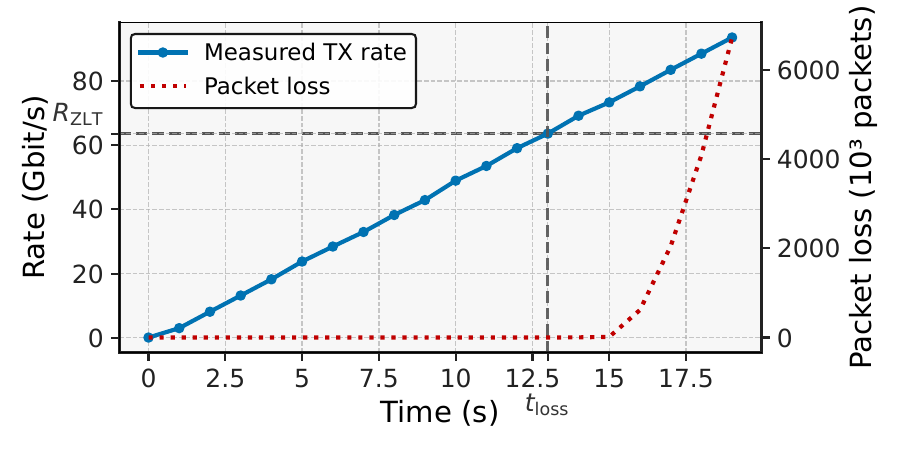}{Zero-loss throughput determination of the receiving application.}
As shown in \fig{pdfs/Baseline}, the application experiences packet loss at $t_{\text{loss}}=\qty{13}{\s}$.
It can therefore process approximately $R_\text{ZLT} = \qty{65}{\gbps}$ without dropping packets.
Any higher rate results in loss.
While this experiment establishes a baseline for the processing capability of the server application, it also showcases the use of the sawtooth pattern to probe for the maximum throughput.

\subsubsection{Packet Loss due to Microbursts}
Second, we test the behavior of the server application under microbursts.
A background \ac{CBR} stream of \qty{20}{\gbps} is generated to emulate steady user traffic.
In addition, a second stream introduces periodic microbursts using a square wave pattern with the parameters $R_{\text{max}}=\qty{80}{\gbps}$, a burst duration of $t = \qty{500}{\us}$, and a period of $T=\qty{10}{ms}$.
Each burst therefore spikes to a combined rate of \qty{100}{\gbps} for \qty{500}{\us} every \qty{10}{ms}.
Traffic is generated for \qty{60}{\s} and the experiment is repeated ten times.
We observed packet loss while the measured 1-second averaged TX rate only showed \qty{24}{\gbps} of the combined \qty{100}{\gbps}.
To visualize the pattern more clearly, we repeat the experiment with longer bursts ($t=\qty{10}{ms}$, $T=\qty{10}{s}$) that are visible on a second scale.
\fig{pdfs/ddos_square} shows the configured (theoretical) rate, the measured TX rate, and the observed packet loss.

\figeps[0.8\columnwidth]{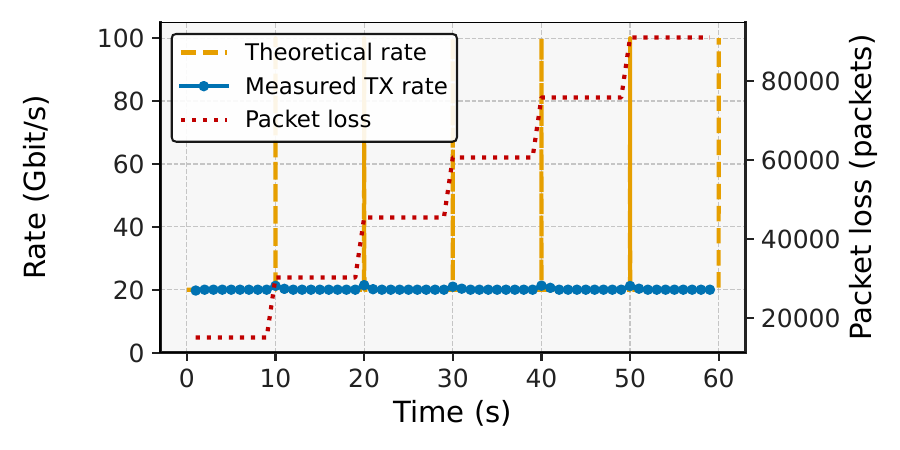}{Configured microburst rate, measured TX rate, and packet loss.}

While the bursts exceed the receiver's \qty{65}{\gbps} saturation point, the 1-second averaged TX rate only shows small fluctuations of up to \qty{21.6}{\gbps}.
This is because the rate counters exposed by P4TG are averaged over \qty{1}{s} intervals, similar to other approaches~\cite{rfc3410, rfc3176}, which smooths out the \qty{10}{\ms} spikes.
As a result, the burst is effectively hidden in the reported rate, and the overload becomes visible only through the packet-loss peaks.

\subsection{Buffer Capacity Measurement and Microburst Impact on Forwarding Device}
\label{sec:eval_buffer}
RFC~8239~\cite{rfc8239} describes a benchmarking methodology for data center networks, including buffer and microburst testing.
We demonstrate an application of pattern shaping for measuring a device's buffer capacity through controlled burst overload, similar to RFC~8239.
We connect P4TG to an Intel Tofino\texttrademark\ 2 switch via a \qty{400}{\gbps} link, returning traffic over a \qty{100}{\gbps} bottleneck link.
P4TG generates square-wave traffic with \qty{64}{B} frames, period $T = \qty{20}{ms}$, and L2 burst rate $R^{L2}_{\text{max}} = \qty{77.24}{\gbps}$ (\qty{101}{\gbps} on L1).
Excess traffic from the burst accumulates in the switch's buffer.
At a critical burst duration $t_{\text{crit}}$, the buffer fills completely and packet loss begins.
We vary the burst duration $t$, i.e., the duration of the high phase of the square wave, to identify $t_{\text{crit}}$ where packet loss first occurs.
The buffer capacity $C$ can then be derived as 
\begin{align}
    C = (R^{L2}_{\text{max}} - R^{L2}_B) \cdot t_{\text{crit}}\label{eq:capacity}
\end{align}
where $R^{L2}_{\text{max}}$ is the L2 burst rate, $R^{L2}_B$ is the L2 bottleneck bandwidth, and $t_{\text{crit}}$ is the shortest burst duration that causes packet loss.
Through binary search, we identify $t_{\text{crit}} \approx \qty{19.6}{\ms}$.
Using \equa{capacity}, with $R^{L2}_{\text{max}} = \qty{77.24}{\gbps}$, \qty{64}{B} frames, and thus, $R^{L2}_B = \qty{76.19}{\gbps}$, we calculate an effective buffer capacity of approximately \qty{2.57}{MB} of frame data.
With \qty{64}{B} frames requiring one \qty{176}{B} cell each in the Tofino\texttrademark\ 2, this results in $\frac{\qty{2.57}{MB}}{\qty{64}{B}} = 40196$ cells.
This value approximately corresponds to the configured buffer size of 40156 cells in the Tofino\texttrademark\ 2.

Next, we leverage the measured buffer capacity to demonstrate a microburst attack on the switch.
With a rate of $R_{\text{max}} = \qty{294}{\gbps}$\footnote{Maximum rate P4TG can generate with \qty{64}{B} frames on a single port.} and the identified buffer size of \qty{2.57}{MB}, the theoretical critical burst duration is approximately \qty{139}{\mu s} according to \equa{capacity}.
In our measurement, we observed packet loss with a minimum microburst duration of $t_{\text{crit}} = \qty{138}{\mu s}$ using a rate of $\qty{294}{\gbps}$, confirming our previous buffer capacity measurement.
Further, while we observed packet loss from buffer exhaustion in the switch, the 1-second rate averaging revealed only \qty{4.44}{\gbps} (1.5\%) of the \qty{294}{\gbps}.
This demonstrates how attackers can cause forwarding-plane buffer exhaustion while remaining nearly invisible to conventional monitoring systems.

\subsection{Impact of Microbursts on TCP Traffic}
\label{sec:eval_tcp}
We demonstrate how P4TG-generated microbursts degrade TCP performance while remaining mostly undetectable by conventional rate monitoring.

\subsubsection{Testbed}
The testbed shown in \fig{pdfs/evaluation_setup_tcp} consists of two servers running an \texttt{iperf3} client and server, P4TG on an Intel Tofino\texttrademark\ 2 switch, and a second Intel Tofino\texttrademark\ 2 for traffic aggregation.

\figeps[0.9\columnwidth]{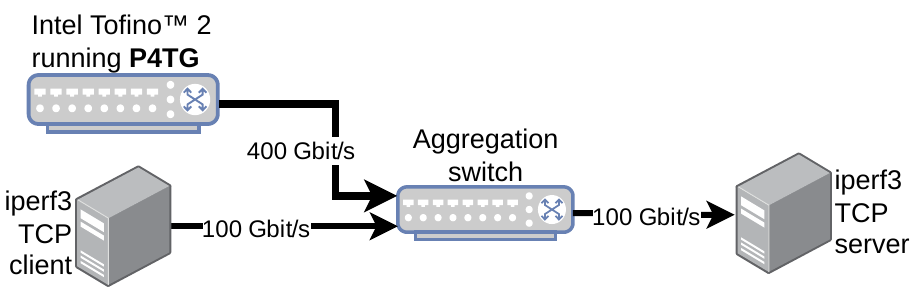}{Testbed for evaluating the impact of microbursts on TCP throughput.}

P4TG is connected via a \qty{400}{\gbps} link and generates UDP microbursts with peak rate $R_{\text{max}}$ and a frame size of \qty{64}{B}.
The generated traffic targets a server running \texttt{iperf3} which receives TCP Cubic traffic from a client.
The aggregation switch combines both traffic streams over a single \qty{100}{\gbps} bottleneck link.
We vary the number of concurrent TCP flows $n$, the period $T$, and the burst duration $t$, denoting each configuration as $T/t$.
Each scenario runs for \qty{60}{s} with ten repetitions.

\subsubsection{TCP Throughput under Microburst Attacks}
\begin{figure}[t]
  \centering
  \begin{subfigure}[t]{0.48\columnwidth}
    \centering
    \includegraphics[width=\linewidth]{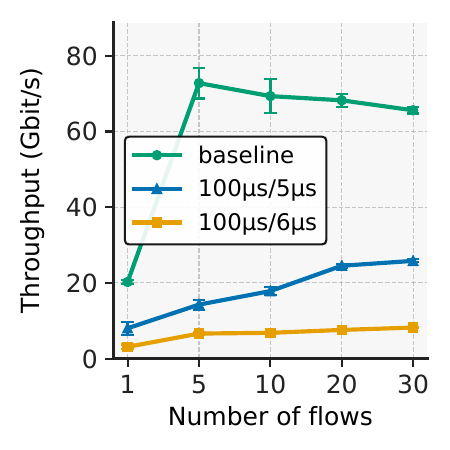}
    \caption{No delay.}
    \label{fig:iperf3_tcp}
  \end{subfigure}
  \begin{subfigure}[t]{0.48\columnwidth}
    \centering
    \includegraphics[width=\linewidth]{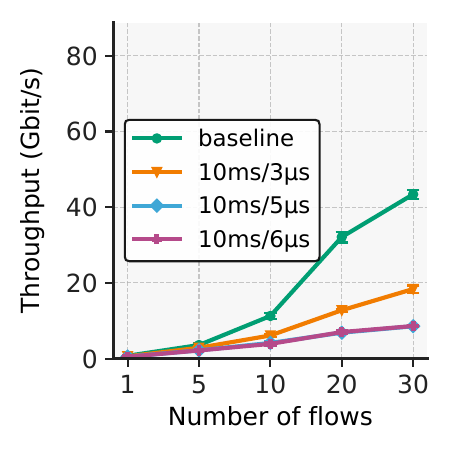}
    \caption{Delay of \qty{10}{\ms}.}
    \label{fig:iperf3_tcp_netem}
  \end{subfigure}
  \caption{Impact of microbursts on TCP traffic. 
  }
  \label{fig:tcp_microbursts}
\end{figure}

\fig{tcp_microbursts} shows the TCP throughput reported by the \texttt{iperf3} server with 95\% confidence intervals.

In \fig{iperf3_tcp}, for $n \ge 5$, the baseline throughput ranges from 60 to \qty{70}{\gbps}.
With microbursts of $R_{\text{max}} = \qty{294}{\gbps}$, and microburst parameters $\qty{100}{\mu s}/\qty{6}{\mu s}$, throughput drops below \qty{8}{\gbps}.
The buffer capacity of the aggregation switch identified in \sect{eval_buffer} saturates approximately at $\qty{139}{\mu s}$ of traffic at an L1 rate of \qty{294}{\gbps}, and at \qty{78}{\mu s} with the TCP baseline of \qty{70}{\gbps} included.
The microbursts are shorter than the switch's buffer capacity, indicating that there is no packet loss on the bottleneck link and the throughput degradation results from server overload rather than a switch buffer overflow.
The 1-s rate averaging measured only \qty{17.6}{\gbps} (9\%) of the \qty{294}{\gbps} for a microburst with \qty{100}{\us}/\qty{6}{\us}, mostly masking the attack.

The testbed's low RTT of approx. \qty{3}{\us} allows TCP to recover quickly from bursts.
To model realistic conditions, we add a constant \qty{10}{ms} delay using \texttt{netem} at the client.
With an increased \ac{RTT}, a single microburst every \qty{10}{ms}, i.e., once per \ac{RTT}, is sufficient to disrupt TCP's congestion control and degrade throughput.
We therefore increase the microburst period to \qty{10}{\ms}.
Without added delay, $T=\qty{10}{ms}$ showed negligible TCP impact.
As shown in \fig{iperf3_tcp_netem}, the baseline drops to at most \qty{45}{\gbps} compared to \fig{iperf3_tcp} due to the added delay.
Further, \fig{iperf3_tcp_netem} shows that the TCP throughput degrades to \qty{8}{\gbps}, similar to the no-delay case with shorter period, while using the same burst durations.
Crucially, rate averaging measured only \qty{0.1}{\gbps} (0.03\%) of the \qty{294}{\gbps} bursts, rendering the attack invisible to conventional monitoring.

\section{Conclusion}
\label{sec:conclusion}
In this work, we extended P4TG with a mechanism for shaping generated traffic with periodic, realistic traffic patterns.
Unlike generators limited to either \ac{CBR} traffic or low generation rates, our design supports fine-grained shaping of periodic traffic patterns, such as sine, sawtooth, square, and flashcrowd, at up to \qty{4}{\tbps}.
Pattern shaping is achieved by control plane sampling and normalization, combined with data plane enforcement at line rate.
While the current implementation targets Intel Tofino\texttrademark\ hardware, the shaping logic uses common P4 constructs, i.e., registers, meters, and \acp{MAT}.
Therefore, the mechanism is portable to other P4-programmable targets, such as the emerging X2 platform of Xsight~\cite{x2}.
Our evaluation demonstrates that the generated traffic closely follows the configured patterns, even at multi-terabit rates.
The scalability analysis quantified the supported pattern periods and table resource requirements, confirming that the approach remains practical across a wide range of configurations.
We further showed that pattern shaping enables practical use cases such as a single-run zero-loss throughput determination using a sawtooth pattern, following the RFC 2544~\cite{rfc2544} methodology, and a buffer capacity measurement with square-wave probing.
Finally, we demonstrated three microburst-based attack scenarios in which short, high-intensity bursts overload a UDP target, a switch buffer, and a shared TCP link while evading detection by monitoring systems with second-scale sampling.
By making periodic pattern shaping practical on programmable switch hardware, this work establishes P4TG as a platform for controlled and repeatable studies of networked systems under \ac{DDoS} attacks, burst-load effects, and time-varying traffic patterns.

\bibliography{bibliography/literature}

\bibliographystyle{unsrt2authabbrvpp}

\end{document}